\documentclass[preprint,12pt]{elsarticle}
\usepackage{cite}
\usepackage{algorithmic}
\usepackage{array}
\usepackage[caption=false,font=normalsize,labelfont=sf,textfont=sf]{subfig}
\usepackage{textcomp}
\usepackage[hidelinks]{hyperref}
\usepackage{amsthm,amsmath,amssymb,amsfonts}
\usepackage{url}
\usepackage[T1]{fontenc}
\usepackage{tikz}
\usepackage[nolist]{acronym}
\usepackage{pgfplots}
\usepackage{caption}
\usepackage{stfloats}
\usepackage{verbatim}
\usepackage{graphicx}
\usepackage{balance}
\pgfplotsset{compat=1.18}

\usepackage{soul}
\soulregister\ac7

\begin{acronym}
  \acro{3-D}{three-dimensional}
  \acro{1G}{first generation}
  \acro{1PPS}{1 pulse per second}
  \acro{2G}{second generation}
  \acro{3G}{third generation}
  \acro{3GPP}{3rd Generation Partnership Project}
  \acro{4G}{fourth-generation}
  \acro{5G}{fifth-generation}
  \acro{5G-NR}{5G new radio}
  \acro{5G-RANGE}{remote area access network for the 5th Generation}
  \acro{5G-DARA}{dynamic spectrum and resource allocation for remote areas}
  \acro{5G-COSORA}{collaborative spectrum Sensing optimized for remote areas}
  \acro{5GCM}{fifth-generation of channel model}
  \acro{5TONIC}{5G Telefonica Open Network Innovation Centre}
  \acro{6G}{sixth-generation}
  \acro{ABG}{alpha, beta and gamma}
  \acro{AC}{alternating current}
  \acro{ADC}{analogue-digital converter}
  \acro{ADR}{angle diversity receiver}
  \acro{AI}{artificial intelligence}
  \acro{AoA}{azimuth of arrival}
  \acro{AoD}{azimuth of departure}
  \acro{ARPU}{average revenue per unit}
  \acro{ARQ}{automatic repeat request}
  \acro{ASIP}{application specific integrated processors}
  \acro{AWGN}{additive white Gaussian noise}
  \acro{B5G}{Beyond 5G networks}
  \acro{BER}{bit-error rate}
  \acro{BLER}{block-error rate}
  \acro{BCH}{Bose-Chaudhuri-Hocquenghem}
  \acro{BRICS}{Brazil-Russia-India-China-South Africa}
  \acro{BS}{base station}
  \acro{CA}{carrier aggregation}
  \acro{CAP}{carrier-less amplitude and phase}
  \acro{CAPEX}{capital expenditure}
  \acro{CC}{constant current}
  \acro{CCR}{constant current reduction}
  \acro{CDF}{cumulative distribution function}
  \acro{CDMA}{code division multiple access}
  \acro{CDS}{conventional diffuse system}
  \acro{CFAR}{constant false alarm rate}
  \acro{CI}{close-in}
  \acro{CIF}{close-in free space propagation path loss with a frequency-weighted exponent}
  \acro{CIR}{channel impulse response}
  \acro{CMMA}{cascaded multi-modulus algorithm}
  \acro{CMOS}{complementary metal-oxide-semiconductor}
  \acro{CoMP}{cooperative multi-point}
  \acro{COST}{European Cooperation in Science and Technology}
  \acro{CP}{cyclic prefix}
  \acro{CQI}{channel quality information}
  \acro{CMOS}{complementary metal-oxide-semiconductor}
  \acro{C-RAN}{centralized radio access network}
  \acro{CR}{cognitive radio}
  \acro{CS}{cyclic suffix}
  \acro{CSI}{channel state information}
  \acro{CSK}{color-shift keying}
  \acro{CSMA}{carrier sense multiple access}
  \acro{CSS}{cooperative spectrum sensing}
  \acro{DC}{direct current}
  \acro{DCO-OFDM}{direct current O-OFDM}
  \acro{DAC}{digital-analogue converter}
  \acro{DAT}{delay adaptation technique}
  \acro{DFT}{discrete Fourier transform}
  \acro{DFT-s-OFDM}{DFT-Spread OFDM}
  \acro{DPD}{digital pre-distortion}
  \acro{DSA}{dynamic spectrum access}
  \acro{DVB}{digital video broadcast}
  \acro{DZT}{discrete Zak transform}
  \acro{ED}{energy detection}
  \acro{EHF}{extremely high frequency}
  \acro{eMBB} {enhanced mobile broadband}
  \acro{EMI}{electromagnetic interference}
  \acro{EoA}{elevation of arrival}
  \acro{EoD}{elevation of departure}
  \acro{EPC}{evolved packet core}
  \acro{eRAC}{enhanced remote area communications}
  \acro{ETSI}{European Telecommunications Standards Institute}
  \acro{EU}{Europe}
  \acro{EVM}{error vector magnitude}
  \acro{FABS}{fast adaptive beam steering}
  \acro{FBMC}{filterbank multicarrier}
  \acro{FDE}{frequency-domain equalization}
  \acro{FDMA}{frequency division multiple access}
  \acro{FD-OQAM-GFDM}{frequency-Domain OQAM-GFDM}
  \acro{FEC}{forward error correction}
  \acro{FoV}{field of view}
  \acro{F-OFDM}{Filtered Orthogonal Frequency Division Multiplexing}
  \acro{FPGA}{field programmable gate array}
  \acro{fps}{frames per second}
  \acro{FSO}{free space optical}
  \acro{FSPL}{Free Space Path Loss}
  \acro{FTN}{Faster Than Nyquist}
  \acro{FT}{Fourier Transform}
  \acro{FSC}{Frequency-Selective Channel}
  \acro{GDB}{Geolocation Database}
  \acro{GFDM}{Generalized Frequency Division Multiplexing}
  \acro{GPP}{General Purpose Processor}
  \acro{GPS}{Global Positioning System}
  \acro{GRCR}{Gershgorin Radii and Centers Ratio}
  \acro{GRE}{Generic Routing Encapsulation}
  \acro{GSCM}{Geometry-based Stochastic Channel Model}
  \acro{GS-GFDM}{Guard-Symbol GFDM}
  \acro{HetNet}{heterogeneous access network}  
  \acro{IARA}{Internet Access for Remote Areas}
  \acro{ICI}{Inter-carrier Interference}
  \acro{ID}{identification}
  \acro{IDFT}{Inverse Discrete Fourier Transform}
  \acro{IDMA}{Interleaved Division Multiple Access}
  \acro{IEEE}{Institute of Electrical and Electronics Engineers}
  \acro{IFI}{Inter-Frame Interference}
  \acro{IFPI}{Interference-Free Pilots Insertion}
  \acro{IMS}{IP Multimedia Subsystem}
  \acro{InH}{indoor hotspot}
  \acro{IoT}{Internet of Things}
  \acro{IP}{Internet Protocol}
  \acro{IPsec}{Internet Protocol Security}
  \acro{IR}{Infrared Radiation}
  \acro{ISI}{inter-symbol interference}
  \acro{ISM}{Industrial Scientific and Medical}
  \acro{ITS}{intelligent transportation system}
  \acro{ITU}{International Telecommunication Union}
  \acro{IUI}{inter-user interference}
  \acro{IVC}{inter-vehicle communication}
  \acro{I2V}{infrastructure-to-vehicle}
  \acro{JEITA}{Japan electronics and information technology industries association}
  \acro{LAA}{license assisted access}
  \acro{LAD}{localization algorithm based on double-thresholding}
  \acro{LBT}{listen-before-talk}
  \acro{LD}{laser diode}
  \acro{LDPC}{low-density parity check}
  \acro{LED}{light emitting diode}
  \acro{LLR}{log-likelihood ratio}
  \acro{LMMSE}{linear minimum mean square error}
  \acro{LoS}{line-of-sight}
  \acro{LPMA}{lattice partition multiple access}
  \acro{LTE}{long-term evolution}
  \acro{LTE-A}{long-term evolution-advanced}
  \acro{M2M}{machine-to-machine}
  \acro{MA}{multiple access}
  \acro{MAC}{medium access control}
  \acro{MANO}{management and orchestration}
  \acro{MAR}{mobile autonomous reporting}
  \acro{MBB}{mobile broadband}
  \acro{m-CAP}{multi-band CAP}
  \acro{METIS}{mobile and wireless communications enablers for the twenty-twenty information society}
  \acro{MF}{matched filter}
  \acro{MFTP}{maximum flickering time period}
  \acro{MIMO}{multiple-input multiple-output}
  \acro{mMIMO}{massive multiple-input multiple-output}
  \acro{mmMAGIC}{millimetre-wave based mobile radio access network for 5G integrated communications}
  \acro{MMSE}{minimum mean square error}
  \acro{mMTC}{massive machine type communications}
  \acro{mmWave}{millimeter wave}
  \acro{MNO}{mobile network operator}
  \acro{MPA}{message passing algorithm}
  \acro{MPC}{multipath components}
  \acro{MPE}{maximum permissible exposure}
  \acro{MPPM}{multiple pulse position modulation}
  \acro{MCS}{modulation coding scheme}
  \acro{MRC}{maximum ratio combiner}
  \acro{MSE}{mean-squared error}
  \acro{MTC}{machine-type communications}
  \acro{MUSA}{multi-user shared access}
  \acro{MVNO}{mobile virtual network operator}
  \acro{N3IWF}{non-3GPP inter-working function}
  \acro{NB-IoT}{narrowband IoT}
  \acro{NEF}{noise enhancement factor}
  \acro{NET}{network}
  \acro{NFV}{network functions virtualization}
  \acro{NGC}{next-generation core}  
  \acro{NIST}{National Institute of Standards and Technology}
  \acro{NLoS}{non-line-of-sight}
  \acro{NOMA}{non-orthogonal multiple access}
  \acro{NR}{new radio}
  \acro{NRZ}{non-return-to-zero}
  \acro{NRZI}{non-return-to-zero-inverted}
  \acro{NYUSIM}{New York University Simulator}
  \acro{O2I}{outdoor-to-indoor}
  \acro{O2O}{outdoor-to-outdoor}
  \acro{OBU}{on-Board Unit}  
  \acro{OM}{operations and maintenance} 
  \acro{OFDM}{orthogonal frequency division multiplexing}
  \acro{OFDMA}{orthogonal frequency division multiplexing access}
  \acro{O-OFDM}{optical orthogonal frequency division multiplexing}  
  \acro{OMA}{orthogonal multiple access}
  \acro{OOB}{out-of-band}
  \acro{OOBE}{out-of-band emission}
  \acro{OOK}{on-off keying}
  \acro{OPEX}{operational expenditure}
  \acro{OQAM}{offset quadrature amplitude modulation}
  \acro{OSNR}{optical signal-to-noise ratio}
  \acro{OTT}{over-the-top}
  \acro{OWC}{optical wireless communication}
  \acro{OWPAN}{optical wireless personal area network}
  \acro{PAPR}{peak-to-average power ratio}
  \acro{PD}{photo-detector}
  \acro{PDF}{probability density function}
  \acro{PDMA}{pattern division multiple access}
  \acro{PDU}{packet data unit}
  \acro{PHY}{physical layer}
  \acro{PLC}{power line communication}
  \acro{PLE}{path loss exponent}
  \acro{PLED}{phosphor-based LED}
  \acro{PLS}{physical layer security}
  \acro{PoC}{proof-of-concept}
  \acro{PSD}{power spectrum density}
  \acro{PSS}{primary synchronization signal}
  \acro{PU}{primary user}
  \acro{PUCCH}{physical uplink control channel}
  \acro{PUSCH}{physical uplink shared channel}
  \acro{PWM}{pulse width modulation}
  \acro{QAM}{quadrature amplitude modulation}
  \acro{QKD}{quantum key distribution}
    \acro{QoE}{quality-of-experience}
  \acro{QoS}{quality-of-service}
  \acro{QPSK}{quadrature phase-shift keying}
  \acro{QuaDriGa}{quasi deterministic radio channel generator}
  \acro{RAN}{radio access network}
  \acro{RAT}{radio access technology}
  \acro{RB}{Resource Block}
  \acro{RBG}{Resource Block Group}
  \acro{RGB}{red-green-blue}
  \acro{RC}{raised cosine}
  \acro{RepC}{repetition coding}
  \acro{RE}{Resource Element}
  \acro{RF}{radio frequency}
  \acro{RLL}{run length limited}
  \acro{RMS}{root mean square}
  \acro{RMSE}{root mean square error}  
  \acro{ROI}{Return on Investment}
  \acro{RPO-OFDM}{reverse polarity O-OFDM}
  \acro{RRC}{Root Raised Cosine}
  \acro{RSA}{Rivest–Shamir–Adleman}
  \acro{RTT}{Round Trip Time}
  \acro{RMIO}{Rural Mobile Infrastructure Operator}
  \acro{RZ}{return-to-zero}
  \acro{SC}{secrecy capacity}
  \acro{SCL}{Successive Cancellation List}
  \acro{SCS}{Subcarrier Spacing}
  \acro{SC-FDE}{Single Carrier Frequency Domain Equalization}
  \acro{SC-FDMA}{Single Carrier Frequency Domain Multiple Access}
  \acro{SCMA}{Sparse Code Multiple Access}
  \acro{SDN}{Software-Defined Network}
  \acro{SDR}{Software-Defined Radio}
  \acro{SDW}{software-defined waveform}
  \acro{SEP}{symbol error probability}
  \acro{SER}{symbol error rate}
  \acro{SIC}{successive interference cancellation}
  \acro{SINR}{signal-to-interference-and-noise ratio}
  \acro{SISO}{single-input single-output}
  \acro{SM}{spatial modulation}
  \acro{SMS}{short message service}
  \acro{SMP}{spatial multiplexing}  
  \acro{SNR}{signal-to-noise ratio}
  \acro{SS}{Spectrum Sensing}
  \acro{SSL}{solid state lighting}
  \acro{STC}{Space Time Code}
  \acro{SSCM}{Statistical Spatial Channel Model}
  \acro{STFT}{Short-Time Fourier Transform}
  \acro{SU}{Secondary User}
  \acro{SUAV}{Small Unmanned Aerial Vehicle}
  \acro{SUMO}{simulation of urban mobility}
  \acro{TAM}{Total Addressable Market}
  \acro{TDMA}{Time Division Multiple Access}
  \acro{TD-OQAM-GFDM}{Time-Domain OQAM-GFDM}
  \acro{TIA}{trans-impedance amplifier}
  \acro{TGI}{temporal ghost imaging}
  \acro{TTI}{time transmission interval}
  \acro{TR-STC}{Time-Reverse Space Time Coding}
  \acro{TR-STC-GFDMA}{Time-Reversal Space-Time Coding Generalized Frequency Division Multiple Access}
  \acro{TVC}{time-variant channel}
  \acro{TVWS}{TV white space}
  \acro{UE} {user equipment}
  \acro{UFMC}{universal filtered multi-carrier}
  \acro{UF-OFDM}{universal filtered orthogonal frequency multiplexing}
  \acro{UHF}{ultra high frequency}
  \acro{UMa}{urban macrocell}
  \acro{UMi}{urban microcell}
  \acro{UPF}{User Plane Function}
  \acro{URLLC}{ultra reliable low latency communications}
  \acro{USRP}{universal software radio peripheral}
  \acro{V2I}{vehicle-to-infrastructure}
  \acro{V2V}{vehicle-to-vehicle}
  \acro{VLC}{visible light communications}
  \acro{VLCC}{VLC consortium}
  \acro{VLP}{visible light positioning}
  \acro{VPE}{vehicle pose estimation}  
  \acro{VHN}{vehicular heterogeneous network}
  \acro{VHF}{very high frequency}
  \acro{VNF}{virtual network function}
  \acro{V-OFDM}{vector OFDM}
  \acro{VOOK}{variable OOK}
  \acro{VPPM}{variable pulse-position modulation}
  \acro{WAVE}{wireless access in vehicular environments}
  \acro{WDM}{wavelength division multiplexing}
  \acro{W-GFDM}{windowed generalized frequency division multiplexing}
  \acro{WHT}{Walsh-Hadamard transform}
  \acro{WIBA}{window-based energy detection}  
  \acro{Wi-Fi}{wireless fidelity}
  \acro{WiMAX}{worldwide interoperability for microwave access}
  \acro{WINNER}{wireless world initiative new radio}
  \acro{WLAN}{wireless local area network}
  \acro{WLE}{widely linear equalizer}
  \acro{WLP}{wide linear processing}
  \acro{WPAN}{wireless personal area network}
  \acro{WRAN}{wireless regional area network}
  \acro{WSN}{wireless sensor networks}
  \acro{ZF}{zero-forcing}
  \acro{ZMCSC}{zero-mean circular symmetric complex Gaussian}
\end{acronym}

\usepackage{adjustbox}

\begin{document}
\sloppy
\begin{frontmatter}
\journal{Physical Communication}

\title{Visible Light Communication for Vehicular Networks: A Tutorial}


\author[1]{Pedro E. G\'oria Silva} \ead{pedro.goria.silva@lut.fi}
\author[3]{Eduardo S. Lima}
\ead{elima@get.inatel.br}
\author[1,4]{Jules M. Moualeu} 
\ead{jules.moualeu@wits.ac.za}
\author[1]{Mohamed Korium}
\ead{mohamed.korium@lut.fi}
\author[1]{Pedro H. J. Nardelli} 
\ead{pedro.nardelli@lut.fi}

\affiliation[1]{organization={Lappeenranta--Lahti University of Technology (LUT)}, addressline={Yliopistonkatu 34}, postcode={53850}, city={Lappeenranta}, country={Finland}}
\affiliation[3]{organization={5G Innovation Office, VS Telecom}, addressline={Lord Cockrane, 616 - Cjs. 601 a 608 - Ipiranga}, postcode={04213-001}, city={São Paulo}, country={Brazil}}
\affiliation[4]{organization={University of the Witwatersrand}, addressline={1 Jan Smuts Ave, Braamfontein}, postcode={2000}, city={Johannesburg}, country={South Africa}}



\begin{abstract}

The advent of the fifth-generation technology promises to bring about more vertical applications and emerging services that include vehicular networks and \acp{ITS}. To achieve their vision of real-time and safety applications, vehicular networks rely on short-range to medium-range communications. One emerging technology that aims to provide reliability and high-data rate in short-range communications is the \ac{VLC}. Due to its remarkable advantages, some studies have recently investigated the integration of \ac{VLC} in vehicular networks and \acp{ITS}. Despite their attractive features, such networks also face several implementation issues. This paper provides an extended tutorial on the implementation of \ac{VLC}-based vehicular networks. To begin with, we present the implementation characteristics of these systems and discuss some related issues. The underlying system considers a general structure with transmitters, channels, and receivers based on photodetectors and cameras, as well as standardization efforts and types of topologies. In addition, we discuss the impact of the sun and artificial light sources, flickering, dimming, throughput enhancement, uplink security, and mobility on practical implementation. Finally, we highlight some key challenges and potential solutions and provide some directions for future research investigations that could constitute an advancement toward the development of commercial \ac{VLC}-based vehicular systems.
		
\end{abstract}


\end{frontmatter}

\acresetall

\renewcommand{\thesection}{\Roman{section}}	


\section{Introduction}
The commercialization of \ac{5G} mobile networks has recently gained momentum throughout the world \citep{8985528}. 
Initially deployed for non-standalone operations such as \ac{LTE} \ac{EPC} networks, the \ac{5G} technology is now adopted
in standalone operations wherein the \ac{NGC} coordinates and controls both the data and
control planes of mobile technologies \citep{giordani2019standalone}. 
As defined by the \ac{ITU}, this technology encompasses four main application areas including \ac{eMBB}, \ac{mMTC}, \ac{URLLC}, and \ac{eRAC} \citep{series2015imt,dias20205g}. Specifically, the \ac{3GPP} has standardized the \ac{5G} \ac{NR} and its Release 15 to focus on \ac{eMBB} applications \citep{ghosh20195g}, while the upcoming Releases are likely to include \ac{URLLC}, \ac{mMTC} and \ac{eRAC} applications. On the other hand, with the emergence of vehicular applications such as \ac{V2V}, \ac{V2I}, and \ac{I2V}, the \ac{ITS} which consists of integrating advanced sensors, connectivity, control and information processing technologies to improve road safety, passenger comfort, traffic flow, and environmental issues, 
has recently gained considerable interest in the research community (e.g., \citep{uysal2015visible,kim2015experimental}). 
Moreover, considerable efforts on \ac{ITS} standardizations have led to the specification of the \ac{WAVE} standard that provides wireless co-existence among transportation services based on the IEEE 802.11p standard~\citep{jiang2008ieee}.

To address the plurality of \ac{5G} features, several studies have explored various emerging technologies which include \ac{HetNet}, \ac{C-RAN} in conjunction with microwave photonics, \ac{mMIMO}, \ac{mmWave} and \ac{OWC} (see \citep{bogale2016massive,lima2020multiband} and the references therein). Among these techniques, \ac{OWC} has emerged as a potential solution to fulfill indoor access and flexibility required for vehicular connectivity in \ac{5G} networks. Typically, \ac{OWC} systems employ a modulated optical beam of ultra-violet, visible, or infrared light, which can propagate through the atmosphere. As a subset of \ac{OWC}, \ac{VLC} 
uses visible light signals to enable high-speed wireless data transmissions. It offers remarkable advantages such as the absence of a licensing requirement and electromagnetic interference immunity, allowing access in areas restricted 
to \ac{RF} and frequency reuse. In addition, \ac{VLC} provides large bandwidths for modulation, improved indoor security, and privacy. Moreover, it is characterized by features of spatial diversity due to a wide detection area of tens of thousands of wavelengths \citep{pathak2015visible}. 

In particular, \ac{VLC} based on \acp{LED} has emerged as a cost-effective, energy-efficient, and secure wireless access technology for addressing \ac{5G}-related challenges and demands for future wireless networks. \ac{LED}-based \ac{VLC} systems are compatible with the \ac{PLC} technology which aims to adopt an electrical network capillarity to transport \ac{RF} signals to multiple \ac{VLC} access points, also known as attocells~\citep{8466585}. Primarily designed for indoor applications (e.g., office, aircraft, homes, hospitals), \ac{VLC} has recently found applicability in underwater and outdoor environments, specifically in \acp{ITS} such as \ac{V2V} and \ac{V2I} \citep{koonen2017indoor,jovicic2013visible,cuailean2017current}.
However, \ac{VLC} faces several implementation drawbacks for both outdoor and indoor scenarios. First, the \ac{LED} bandwidth available for modulation is quite limited, and the current commercial \acp{LED} 3-dB bandwidth is only a 
few MHz \citep{chi2018led}. Therefore, it is necessary to enhance the \ac{LED} bandwidth and consequently the throughput performance. Second, \ac{VLC} systems are severely degraded by sunlight and artificial light sources, which increase the noise at the receiver and could saturate the photodetector at high powers \citep{alsulami2018optical}. Third, the emitting light power is also a concern. In this regard, \ac{VLC} systems could avoid high optical powers and provide a safe environment for the eyes \citep{OWreport}. Moreover, depending on the modulation used, indoor \ac{VLC} is prone to flickering which is the fluctuation in light intensity or brightness that is perceptible to the human eyes, causing discomfort and health risks \citep{wilkins2010led}. 

\ac{VLC} has recently gained considerable attention in the research community thanks to its promising features. In \citep{vappangi2019concurrent}, the authors have proposed a \ac{VLC} system with modulation techniques based on \ac{OFDM} and other modulation formats that include \ac{CAP} and \ac{m-CAP} for enhancing both the power and computational efficiencies. In addition, multiple access techniques have been integrated in \ac{VLC} systems. In \citep{karunatilaka2015led},
Karunatilaka \textit{et al.} have provided a comprehensive survey on \ac{VLC} for indoor applications where the advantages of \acp{LED} compared to traditional lighting technologies are highlighted. Moreover, a detailed discussion on modulation schemes and dimming techniques for indoor \ac{VLC} have been provided, and some approaches that can improve the performance of \ac{VLC} systems, have been presented. The potential applications and limitations of the \ac{VLC} technology have been elaborated.
In \citep{cuailean2017current}, Căilean and Dimian
have reported on the challenges related to VLC usage in outdoor applications such as automotive communications. The authors have also argued that the full potential of \ac{VLC} can be achieved by addressing and solving the challenges aforementioned, further enabling the usage of \ac{VLC} in transportation applications. Lastly, the authors have provided some future research directions that could potentially make \ac{VLC} a reliable vehicular communication technology. 
In \citep{memedi2021vehicular}, the authors have revisited the state-of-the-art in \ac{VLC}-based vehicular systems with special emphasis on the \ac{VLC} channel. Moreover, open issues and challenges have been identified in this survey, which serves as a guide to the relevant literature and a reference for beginners and experts in the field, respectively. 
In \citep{zhuang2018survey}, Zhuang \textit{et al.} have provided a comprehensive review of a novel \ac{LED} positioning technology, wherein both the characteristics and principles of the underlying system are thoroughly discussed. In addition, a classification of outdoor \ac{VLC} positioning applications has been reviewed. 

However, the works in \citep{vappangi2019concurrent,karunatilaka2015led,cuailean2017current,memedi2021vehicular,zhuang2018survey} seldom address mobility in both indoor and outdoor (specifically pertaining to vehicular networks) \ac{VLC} systems, since it depends on \ac{LoS} communications. In this regard, the implementation of \ac{VLC} in vehicular networks where the receiver movement, \ac{NLoS} conditions and shadowing induced by humans or objects are taken into account, presents novel challenges Such challenges serve as a motivating factor for this tutorial paper. 

The present work provides a comprehensive overview of the \ac{VLC} focusing on \ac{5G} indoor and vehicle communications. Section \ref{Syst_char} provides a comprehensive overview of the general \ac{VLC} structure focusing on the transmitter, receiver, \ac{LoS} and \ac{NLoS} channels and standardization efforts and topology. Section \ref{challenges}
presents the \ac{VLC} usage for \ac{5G} indoor scenarios, and discuss the associated implementation issues as well as the solutions to overcome such issues, while Section \ref{concl} presents some concluding remarks and highlights some future research studies. 
\section{VLC System Characteristics}
\label{Syst_char}
As aforementioned, \ac{VLC} has recently been adopted as a promising alternative to the \ac{RF} technology. This is due to the fact that the visible light spectrum carrier in the visible light frequency (380-780 nm) enables a bandwidth of up to 1000 times that of \ac{RF} communications~\citep{cuailean2017current}.
%
%
Moreover, the non-regulation of the visible light spectrum directly yields a low-cost implementation of the \ac{VLC} technology and requires a reduced area in comparison to the \ac{RF} technology. 
A notable advantage of \ac{VLC} is the high data rates which achieve up to $10$ Gb/s.\citep{hussein201520,Chun7454689}. 
Although the practical applications and integrated solutions to current communication systems, such as 5G among others, for \ac{VLC} systems are still in their infancy, some impressive results have made this technology a promising candidate for future communication systems.

\begin{figure*}[h]
 	\centering
 	\includegraphics[width=0.95\linewidth]{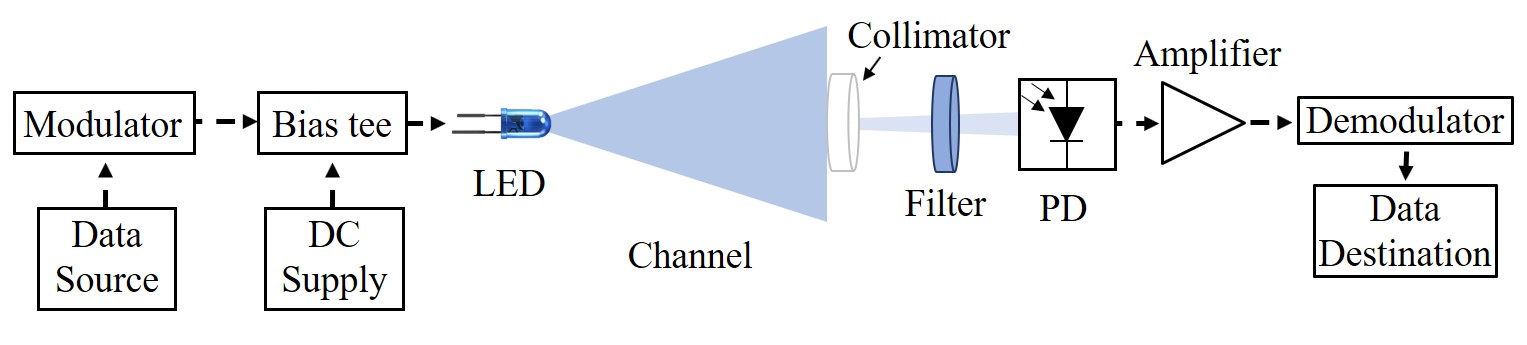}
 	\caption{Visible  light  communication system block diagram.}
	\label{fig: esquema}
 \end{figure*}

In light of the above, it is imperative to discuss the block diagram of a \ac{VLC} system. 
In this regard, Fig. \ref{fig: esquema} illustrates the main components of a \ac{VLC} system. 

As the name suggests, the communication takes place between transmitter and receiver blocks \footnote{In \ac{VLC} systems, a \ac{LoS} is generally present except when there is obstacle that prevents such a direct communication.} is done through the modulation of data in electrical pulses via the \ac{LED}.
These electrical pulses are then converted into electromagnetic waves operating at light frequencies visible by the \ac{LED}, and subsequently received by the \ac{PD} or by a receiver equipped with a video camera.
The \ac{PD} or the video camera converts light beams into electrical signals, which are then filtered and demodulated to recover the transmitted information.
Other components shown in Fig. \ref{fig: esquema} are the \ac{DC} supply and collimator which is a device used to narrow a beam of particles or waves. 
Although they are not essential in the operation of a \ac{VLC} system, they are often implemented in certain scenarios (e.g., wherein ambient light is used).
In the remainder of this section, we discuss in detail the main components of the \ac{VLC} system, i.e., transmitter and receiver structures, \ac{VLC} system design, channel models, and standardization efforts.

\subsection{VLC Transmitter}

In a \ac{VLC} system, the transmitter modulates the information signal in a light beam required by the corresponding propagation medium.
This can be done by combining the data with the \ac{DC} voltage before feeding the \ac{LED}.
In general, \ac{VLC} allows modulation schemes ranging from simpler modulations, such as \ac{OOK}, to more sophisticated ones, such as \ac{OFDM}-\ac{QAM} \citep{Yeh2017}.
Hence, the choice of the modulation scheme in \ac{VLC} systems is crucial and should be carefully evaluated before its adoption.
This is due to the fact that sophisticated modulation schemes may require additional hardware costs in \ac{FPGA} and therefore, increase the costs associated with the implementation a \ac{VLC} transmitter.

Moreover, the implementation of \ac{VLC} systems is limited due to the intricate characteristics of commercialized \acp{LED} in the simultaneous provision of data and lighting.
Therefore, it is essential that the flicker caused by the modulation of the light beam is imperceptible to the human eye, and is jointly harmless to the overall human's wellbeing.
More precisely, the flicker speed of the \ac{LED} is a determinant factor of the bandwidth for the \ac{VLC} system.
However, the flickering phenomenon does not change the average intensity of the ambient light.
In other words, the \ac{VLC} system does not change the lighting pattern of the environment, which represents another feature imposed on the \ac{VLC} transmitter.
Consequently, the standard light intensity of the environment defines a crucial parameter of the \ac{VLC} system design, the transmitter protection.

The response speed of the \ac{LED} is directly proportional to the amount of information per unit of time that the \ac{VLC} system is capable of transmitting. 
The illumination pattern and \ac{LED} power restrict the coverage area of the \ac{VLC} system. 
With rapid advances in technology, the \ac{SSL} industry is now capable of producing \acp{LED} with a switching frequency of only a few tens of MHz. 
In contrast, the data transfer speed for \ac{VLC} systems are in the order of multi-Gb/s. As a result, the slow \acp{LED} represent a potential bottleneck for future \ac{VLC} implementations~\citep{Zafar7842427}.
With this in mind, it has been observed that the current consumer trend aims to replace incandescent, fluorescent and halogen lighting with \ac{LED} lamps.
The popularity of the \ac{LED} market is motivated by that, any \ac{LED} light source can be transformed into a \ac{VLC} emitter, and can therefore ensure three key characteristics: energy efficiency, sustainability, and reliability.

\subsection{VLC Receiver}
In order to demodulate the information sent from a light source, the receiver must absorb the light energy before converting it into an information suitable to the end-user.
Initially, the light beam should be captured and converted into an electrical signal. This process is carried out by a \ac{PD} or a video camera.
After filtering and amplifying the electrical signal, it is then demodulated and decoded, depending on the modulation scheme, by an embedded microcontroller-based module or \ac{FPGA}. 
An important parameter used TO assess the network coverage and data reception is the receiver \ac{FoV}. A narrow \ac{FoV} yields a low amount of interfering signal absorbed by the receiver and a reduced coverage area. Conversely, the receiver will inevitably be subject to significant noise due a large coverage stemming from a wide \ac{FoV}.

The application of the \ac{PD} as an optical receiver is more common than video cameras in \ac{VLC} systems. 
These photosensitive elements have a considerably large bandwidth and, consequently, make receiving data at high rates possible at the receiver.
On the other hand, the \ac{PD} has a peculiar disadvantage in that the level of interference present in a receiving \ac{PD} is significant.
This is because unwanted signals present outside the band of interest are received by the \ac{PD}. 
To address this issue, optical filters capable of rejecting unwanted spectral components can be adopted.
These filters can block specific spectral segments, such as \ac{IR} components, or allow the passage of a narrow frequency band. 
If the transmitter operates with a white \ac{LED} and the \ac{VLC} system works with a high data rate, it is advisable to apply narrow-band filters corresponding to the frequencies surrounding the blue colour at the receiver.
The reason for this is that a combination of blue \ac{LED} and yellow phosphor produces white light and, in this case, only the blue colour is relevant for signal processing.
There is a considerable improvement in the performance of a \ac{VLC} receiver when optical filters are employed.

Another approach to convert the light beam into an electrical signal is through the use of a video camera or an image sensor of a \ac{PD}. 
The motivation for this approach is supported by its practicality. In other words, the vast majority of modern mobile equipment is manufactured with an integrated camera (e.g. smartphones, tablet and laptops).
The image sensor consists of a high number of \ac{PD} arranged in a matrix or an integrated circuit. 
Despite its attractive features as previously mentioned, the use of a video camera or image sensor in \ac{VLC} systems have some drawbacks. 
For instance, the image sensor has a lower noise performance than one independent photo-element.
In addition, the number of \ac{fps} in a video camera, which does not exceed hundreds of \ac{fps}\footnote{It is common to find video cameras on the market with a rate of 50 \ac{fps}.}, limits the maximum data transmission rate.
In a simplified analysis, each frame can be associated as a sample of the transmitted signal.
Consequently, the reception band would be limited in the best-case scenario to 100 Hz, according to Nyquist's criterion.
Thus, it is evident to quantify the limitation of a \ac{VLC} system when using a video camera with low \ac{fps}. 
A procedure known as rolling shutter which consists of reading line-by-line pixels instead of reading the entire matrix at once can be used to improve the data rate of video cameras up to several kb/s, and that is still considerably below what is expected of a \ac{VLC} system~\citep{Ji7023142}.

\subsection{Channel Model}
The radiation pattern of the transmitter and the receiver \ac{FoV} define a characteristic of the \ac{VLC} channel: the \ac{LoS} between the transmitter and receiver. 
It is defined as a signal emitted by the transmitter that arrives at the receiver without experiencing any reflection, refraction, or shadowing.
However, if the received signal contains at least one reflected one, it is referred to as \acl{NLoS}.
A \ac{VLC} system with \ac{LoS} between the transmitter and the receiver is deemed to be more robust to unwanted effects of multipath, primarily if the receiver \ac{FoV} and the opening beam of the transmitter are narrow.
However, the existence of at least one direct path between the transmitter and the receiver yields a \ac{VLC} system susceptible to the shadowing effect~\citep{Kahn554222}.
On the other hand, \ac{NLoS} paths stem from reflections in the surroundings, such as walls, ceilings, and other objects. 

We can classify the \ac{LoS} and \ac{NLoS}, into direct, indirect, and hybrid links. The hybrid scenario represents a combination of direct and indirect links.
When mobility is required in indoor configurations, the best transmission is through the indirect path.
In this way, the transmitter can cover a larger area at the expense of signal dispersion, which will severely impair the maximum data transmission rate. 
Moreover, due to the larger area, the energy at the receiver will be less than that collected using a direct link.

Commonly used in the literature, the \ac{CDS} employs a wide receiver \ac{FoV} and a transmitter with a broad radiation pattern~\citep{Kahn554222}. 
In this technique, the receiver \ac{FoV} and the light beam from the transmitter point to the same reflective object. 
As a result, the signal is received via multiple reflections.
When using specific receivers, such as the triangular pyramidal fly-eye diversity receiver, the \ac{ISI} is mitigated in indoor environments since the value of the receiver \ac{FoV} has been optimized~\citep{Ghamdi1256749}. 
An indoor optical wireless channel using intensity modulation with direct detection~\citep{Kahn554222} can be fully characterized by its impulse response $h(t)$. 
In addition, the instantaneous current received in the \ac{PD} at a certain position due to $M$ reflecting elements is given by~\citep{Ghamdi1264191}
\begin{align}
    y(t, \text{Az}, \text{El}) = \sum_{m=1}^{M} R \times x(t) \otimes h_m(t, \text{Az}, \text{El})+N,
\end{align}
where $t$ is the absolute time, Az and El are the directions of arrival in azimuth and elevation, respectively, $x(t)$ is the instantaneous optical power of the transmitter, $R$ represents the receiver responsivity, $N$ and is the background noise, and $\otimes$ denotes the convolution operation.

Another relevant characterization parameter of an optical wireless channel is the delay spread, which can be obtained through simulations or field measurements.
The \ac{RMS} delay spread is an excellent way to measure the delay spread of an impulsed response which is given by~\citep{Ghamdi1264191}
\begin{align}
    D=\sqrt{\frac{\sum (t_i-\mu)^2 P_{r_i}^2 }{\sum P_{r_i}^2}},
\end{align}
where $t_i$ is the delay time associated with the received optical power $P_{r_i}$ ($P_{r_i}$ reflects the impulse response $h(t)$ behavior), and $\mu$ is the mean delay given by
\begin{align}
    \mu=\frac{\sum t_i P_{r_i}^2}{\sum P_{r_i}^2}.
\end{align}
%
\subsection{Standardisation Efforts}
Efforts to standardize the \ac{VLC} technology was initiated in 2003 by a \ac{VLCC}.
In order to accelerate the deployment and commercialization of the \ac{VLC} technology, the \ac{VLCC} proposed two standards in 2007 namely: JEITA CP-1221 for a \ac{VLC} system and JEITA CP-1222 for a \ac{VLC}ac{ID} system that was later accepted by \ac{JEITA}.
However, both standards showed some limitations in terms of data transmission rate.
In the best case scenario, a rate of $4.8$ kbps is expected given a low value for modern requirements.

It is not until 2009 that \ac{VLCC} introduced the first \ac{IEEE} 802.15.7 standard. 
Some revisions were subsequently made, the latest being in 2018~\citep{8697198}.
\ac{IEEE} 802.15.7 establishes patterns for local and metropolitan networks in which \ac{OWC} is used to cover a short range. 
The media must be optically transparent and the light wavelength must range from $190$ nm to $10,000$ nm.
The standard defines a \ac{PHY} sublayer and \ac{MAC} sublayer. Concerning the maximum data transfer rate, this IEEE standard is capable of providing up to $96$ Mbps, which is sufficient to support audio and video multimedia services. 
In addition, IEEE 802.15.7 also features some scenarios such as optical link mobility, compatibility with various light-providing infrastructures, mitigation of light interference due to noise from sources such as ambient light, and a \ac{MAC} sub-layer that suits the exceptional needs of visible links.
It also deals with optical communications involving camera-based receivers, that is, communications in which the receivers are digital cameras with a lens and image sensor.
An important aspect is the safety of the human eye health which is also regulated by this standard.
The main benefits provided by IEEE 802.15.7 are the use of an unlicensed spectrum of hundreds of terahertz, immunity to electromagnetic interference and non-interference with \ac{RF} systems, additional security, and communication augmenting and complementing existing services.

The IEEE 802.15.7 standard classifies the devices used in \ac{VLC} systems within three categories: infrastructure, mobile, and vehicles.
The infrastructure class refers to all potential objects belonging to the infrastructure that will be used by the \ac{VLC} system. 
Any end device on the network that can move freely, such as a cell phone or notebook, is classified as mobile by the IEEE 802.15.7 standard.
Cars, trucks and other means of transport are in the vehicle class. Table I presents further details on the classes and their characteristics \citep{8697198}.
\begin{table}[h]
\centering
\caption{Device classification, adapted from~\citep{8697198}}
\label{tab: Device classification}
\begin{tabular}{|l|c|c|c|}
\hline
                  & \multicolumn{1}{l|}{Infrastructure} & \multicolumn{1}{l|}{Mobile} & \multicolumn{1}{l|}{Vehicle} \\ \hline
Fixed coordinator & Yes                                 & No                          & No                           \\ \hline
Power supply      & Ample                               & Limited                     & Moderate                     \\ \hline
Form factor       & Unconstrained                       & Constrained                 & Unconstrained                \\ \hline
Light source      & Intense                             & Weak                        & Intense                      \\ \hline
Physical mobility & No                                  & Yes                         & Yes                          \\ \hline
Range             & Short/long                          & Short                       & Long                         \\ \hline
Data rates        & High/low                            & High                        & Low                          \\ \hline
\end{tabular}
\end{table}

The work of Wang \textit{et al.} for an OpenVLC system provides an interface between \ac{VLC} front-end and the embedded Linux platform~\citep{Wang2014}.
This OpenVLC system is flexible and rapid prototyping, and is open source.
Furthermore, it shares some characteristics with IEEE 802.15.7 and proposes the use of a few, such as a time-division duplex. 
However, the maximum data transmission rate for the last release (openVLC 1.3~\citep{Galisteo8767252}) is only 400 kbps, noticeably lower than that of IEEE 802.15.7.
%

%
\subsection{Topologies}
\begin{figure}[ht]
 	\centering
 	\includegraphics[width=1\linewidth]{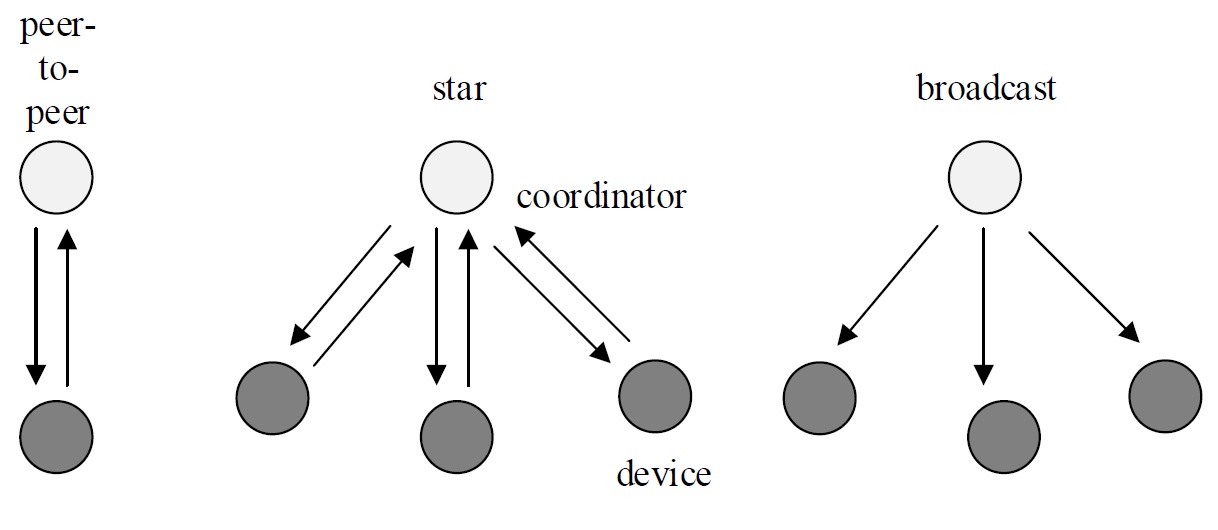}
 	\caption{MAC topologies, adapted from~\citep{8697198}.}
	\label{fig: topologies}
 \end{figure}
\Ac{VLC} systems can operate in different topologies.
However, IEEE 802.15.7 defines peer-to-peer, star, or broadcast topology as standard in \ac{VLC} networks, as shown in Figure~\ref{fig: topologies}. 
First, in a peer-to-peer topology, one of the two devices assumes the role of coordinator.
Moreover, it provides communication between two devices as long as both are in the same coverage area.
One of the pairs in a peer-to-peer \ac{VLC} must assume the role of coordinator. This selection of the coordinator can be made based on the device that first communicates on the channel.
Second, when several devices communicate exclusively with a central device, termed a coordinator, it is referred to as star topology. 
There is no interdependence between two \ac{VLC} networks organized in a star topology, i.e., a star network operates autonomously from all other star networks.
The choice of exclusive \ac{OWPAN} identifiers for each network within the same coverage area enables independent operation of star topologies.
Last but not least, the formation of a specific network is unnecessary in a broadcast topology, that is, the destination address is not required. 
In this device setup, there is a transmitter that disseminates information throughout the coverage area with a unidirectional communication in a broadcast topology.

\section{VLC Challenges in Indoor 5G and Vehicle Applications}
\label{challenges}

The \ac{VLC} technology have emerged as a potential candidate to enable Gbit/s throughput in \ac{5G} and beyond indoor environments and to ensure reliable communication among vehicles.
Despite its remarkable and attractive features, the \ac{VLC} technology faces some implementation hurdles that can hinder the deployment and commercialization of \ac{VLC}-based systems. 
This section describes some implementation challenges pertaining to the \ac{VLC} technology, as well as potential solutions to overcome the above issues and research directions.

\subsection{Influence of Artificial and Natural 
Light Sources}
 
\begin{figure}[ht]
 	\centering
 	\includegraphics[width=1\linewidth]{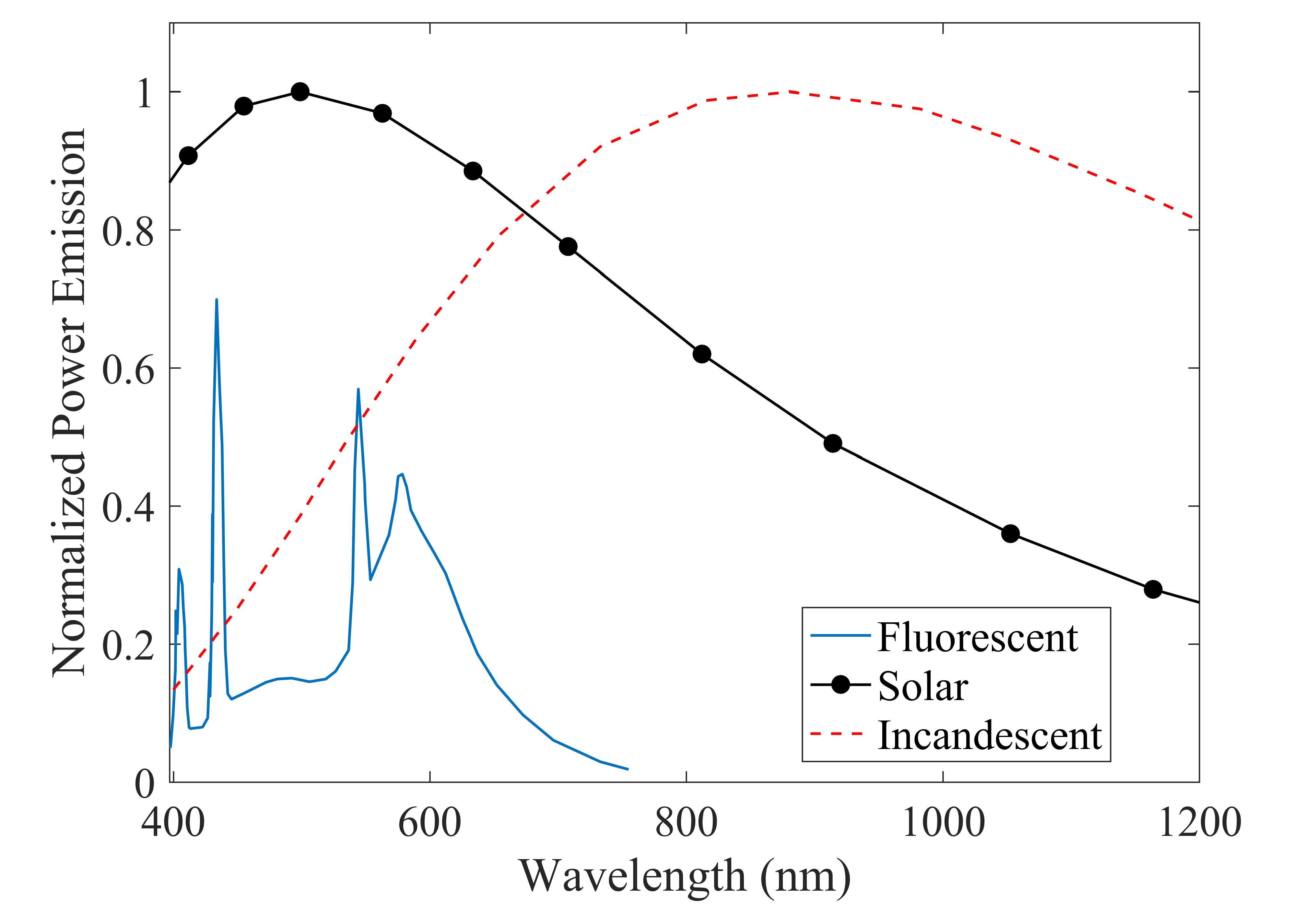}
 	\caption{Spectral density distribution of incandescent and fluorescent lamps, and the emissions of the solar radiation as a function wavelength.}
	\label{Emission}
 \end{figure}

The light emanating from artificial light sources such as fluorescent, incandescent, and neon lamps causes interference and noise to the received optical signal and can critically degrade the performance of a \ac{VLC} system  \citep{alsulami2018optical}. 
All of these light sources emit a substantial amount of power in the wavelength ranging from visible to near-infrared, which is the same wavelength band as that of the \ac{VLC}. 
Such an undesirable emission increases the total optical power at the receiver, which can saturate or even damage the \ac{PD}, besides adding shot-noise \citep{boucouvalas1996indoor}.
Fig.~\Ref{Emission} illustrates the spectral density distribution of incandescent and fluorescent lamps, and the emission of solar radiation emission as a function of wavelength. 

One can note that the sunlight emits in the entire frequency range with emission peak around 500 nm. 
This unmodulated power at the receiver is converted to shot noise, degrading the overall \ac{VLC} performance. 
Incandescent lamps also report power emission in the entire wavelength range. Such lamps are fed and modulated by the \ac{AC} supply (operating at 50 or 60~Hz). 
After the photodetection process, an electrical signal is generated with a bandwidth less than 2 kHz. 
In \citep{moreira1997optical}, Moreira \textit{et al.} have tested six different types of incandescent lamps in terms of interference. 
The results obtained in \citep{moreira1997optical} showed an analogous response for all lamps, that is, a peak at $100$ Hz and harmonics up to 2~kHz.
It is worth mentioning that the harmonics higher than 800~Hz presented components 60~dB below the fundamental harmonics. 
On the other hand, the fluorescent lamp does not emit over the entire range and presents peaks around 450~nm, 550~nm and 580~nm. 
After the photodetection process, the electrical signal presents a total bandwidth of approximately 20~kHz \citep{borah2012review}. 
In addition, the detected signal depends on the electrical signal driven by conventional ballasts or electronic ballasts. 
The former (conventional ballasts) present substantial interference signal up to several kHz, while the latter (electronic ballasts) result in a lower emission power with a response higher than 1~MHz \citep{moreira1997optical}.     

Several techniques have been proposed to overcome these performance penalties. Since the noise current directly depends on the optical power at the \ac{PD} input and the \ac{PD} bandwidth, electrical and optical passband filters can significantly reduce the shot noise \citep{borah2012review}. 
In the case of a fluorescent lamp, positioning the \ac{VLC} emission wavelength outside of the lamp emission and using an optical filter mitigates a large part of the interference and noise.
In this regard, Chang \textit{et al.} have
proposed a filter-based array for \ac{WDM}, containing multiple filters with different
bandpass responses. 
The results demonstrated that the use of the filter enables one to achieve a higher \ac{SINR} ~\citep{chang2011interference}. 
In \citep{sindhubala2017simulation}, the performance analysis of the \ac{RZ}-\ac{OOK} \ac{VLC} system was investigated assuming some
optical background noise. 
Optical and electrical filters have been used to evaluate system performance in terms of \ac{BER}, eye opening, and distance between transmitter and receiver.

The coding or modulation scheme employed in the \ac{VLC} system also plays an important role in reducing the background optical interference.
Liu \textit{et al.} demonstrated a reduction in the background optical interference within the low-frequency band using the \ac{NRZI} code \citep{liu2013employing}. 
In addition, a performance comparison between \ac{NRZ} and \ac{NRZI} codes in a \ac{VLC} system was provided. 
Chow \textit{et al.} studied the Manchester coding in an effort to mitigate the background optical noise without providing a theoretical and numerical analysis \citep{6409954}. 
The experimental results revealed that Manchester coding effectively minimizes the noise generated by \ac{AC}-\acp{LED} and fluorescent light.

\subsection{Flickering}

Widely observed in \ac{VLC}, flickering represents the fluctuation in light intensity or brightness that is perceptible by the human eye to the extent of causing discomfort and serious health risks \citep{wilkins2010led}.
These health risks can be immediate, causing epileptic seizures, or can be a consequence of eye exposure, such as headaches, malaise, and blurred vision. 
The immediate effects are due to visible flickering, which ranges from 3 to 70~Hz while the latter is associated with higher flicker frequencies and is not perceptible by the human eye \citep{wilkins2010led}.
Furthermore, flickering is noticeable depending on the modulation scheme, flicker frequency, code, data rate, brightness, just to name a few \citep{vappangi2019concurrent}.
A \ac{VLC} system must operate at the \ac{MFTP}, which is defined as the maximum time period for which the intensity or brightness of the light can fluctuate without being noticed by the human eye. 
It then follows that a flicker frequency higher than 200~Hz or an \ac{MFTP} lower than 5~ms ensures a safe \ac{VLC} environment \citep{wilkins2010led}.

Since \ac{VLC} systems aim to simultaneously provide lighting and data transmission, mitigating the flicker is also an intricate task.
To address this issue, the IEEE 802.15.7 standard defines some methods to mitigate both intra-frame and inter-frame flickering by employing the \ac{RLL} codes, namely Manchester, 4B6B or 8B10
\citep{hranilovic2004short}. 
In addition, the standard recommends \ac{CSK}, \ac{OOK} and \ac{VPPM} for throughput up to 96~Mbit/s \citep{hranilovic2004short}. 
Thummaluri \textit{et al.} investigated low-complexity encoding and decoding algorithms based on high-rate \ac{RLL} codes and \ac{MFTP} to mitigate the flickering effect \citep{8719334}. 
A performance comparison in terms of code rate, \ac{PAPR} and \ac{BER} between the \ac{RLL} codes and existing ones, was presented. 
In addition, constant-envelope \ac{OFDM} was investigated in \ac{VLC} to regulate the flickering requirements. 
In \citep{vd2019adaptation}, Zwaag \textit{et al.} experimentally showed that the \ac{PAPR} reduction can be used to accomplish the flickering requirements in \ac{LoS} \ac{SISO} \ac{VLC} systems. 
However, this approach can reduce the overall performance of the \ac{VLC} due to the transmission of low-\ac{PAPR} signals.

\subsection{Dimming}

Since {VLC} focuses on simultaneously providing lighting and communication, dimming techniques can be adopted to enable an energy-efficient environment through intelligent lighting solutions. 
To this end, efficient dimming control mechanisms are investigated in an effort to ensure a trade-off between lighting and communication, since traditional methods directly impact \ac{VLC} performance, reducing link throughput~\citep{kumar2019comprehensive}. 
Typically, there are two dimming approaches applied to \acp{LED}: analog dimming and digital dimming \citep{karunatilaka2015led}. 
In the analog technique, the \ac{LED} brightness is configured by adjusting the current of the \ac{LED} driver current, since its emission is directly proportional to the \ac{CC} level. 
This method is also known as \ac{CCR} and has been considered a cost-effective dimming approach for experimental \ac{VLC} systems. 
The IEEE 802.15.7 standard investigates the effect of varying the current for each light source on the analog \ac{CSK}-based dimming  \citep{hranilovic2004short}. 
In \citep{de2020rgb}, Oliveira \textit{et al.} studied an \ac{RGB}-based indoor \ac{VLC} system with the use of the \ac{5G} \ac{NR} standard. 
Moreover, the authors evaluated the performance of the \ac{VLC}-based system in terms of the \ac{RMS} \ac{EVM} as a function of the bias current for each color. 
Numerical results demonstrated that 3GPP requirements for a 10-MHz \ac{5G} signal at 2-m link reach could be achieved for a bias current higher than 240~mA.  

In a digital setup, a digital pulse train drives the \ac{LED}, and thus, the average duty cycle can impact the \ac{LED} dimming. 
Several modulation schemes that maintain the average duty cycle have been used to control the \ac{LED} dimming \citep{karunatilaka2015led}. 
In \citep{4138184}, Sugiyama \textit{et al.} studied brightness control methods for \ac{VLC} systems based on the depth of \ac{PWM} and pulse modulation. 
In \citep{lee2011modulations}, the authors proposed \ac{VLC} modulation schemes that provide dimming control such as \ac{VOOK}, \ac{MPPM}, \ac{VPPM}. 
In \citep{elgala2013reverse}, Elgala and Little proposed an \ac{O-OFDM} as a promising modulation technique for \ac{VLC} dimming . 
A \ac{RPO-OFDM} was investigated to combine both the \ac{O-OFDM} and the \ac{PWM} dimming signals, with the aim of balancing between the radiated optical flux and the perception of the human eye. 

Since \ac{VLC} system relies on lighting for data transmission, the absence of light remains a concern to be addressed.
A potential solution is to employ a hybrid \ac{VLC} system including either \ac{RF} or \ac{mmWave} links, that could yield significant gains in throughput with zero or low \ac{LED} emission (acceptable to the human eye).
In \citep{6162564}, Borogovac \textit{et al.} investigated the lighting limit defined by users as off-state to achieve robust data coverage by using low-complexity \ac{VLC} devices. 

\subsection{Throughput Enhancement}

It is expected that \ac{VLC} systems can provide optical-fiber-like data rates to meet \ac{5G} and beyond requirements, given the large available bandwidth in the visible spectrum(\~300 THz) bandwidth available. 
Despite the huge visible spectrum, the 3-dB bandwidth of commercial \acp{LED} is limited, reaching only a few MHz \citep{chi2018led,grubor2008broadband}.
In recent years, various studies have investigated techniques to improve the throughput performance in \ac{VLC} systems. 
Moreover, several techniques have been proposed to overcome bandwidth limitations, as aforementioned. 
A simple and low-cost approach is to apply a blue optical filter to the receiver to enhance the 3-dB bandwidth. However, the system throughput remains low compared to the bandwidth \citep{karunatilaka2015led}. 
Laser diodes have also been employed as a \ac{VLC} transmitter, enabling higher bandwidths. In \citep{9187349}, Wei \textit{et al.} studied a 6.9 Gbit/s \ac{VLC} system with functional transmission distance based on a white-light phosphor laser.

In addition, several works have proposed techniques to linearize and compensate the \ac{LED} frequency response using a pre-distorter or an equalizer. 
In this regard, Sheu \textit{et al.} employed a pre-distorter at the transmitter to linearize the \ac{LED} through a \ac{DCO-OFDM} \citep{8093774}. 
Unlike a pre-distorter which can be placed at the transmitter, an equalizer circuit can be implemented at the transmitter, receiver or in a hybrid approach. 
Huang \textit{et al.} proposed a cascaded amplitude equalizer followed by a blue filter \citep{huang20151}. This resulted in an improvement in the LED bandwidth from 17~MHz to 366~MHz and attained 1.6~Gbit/s exploiting 16 \ac{QAM}-\ac{OFDM} over a 1-m link.
In \citep{7296576}, Wang \textit{et al.} experimentally demonstrated the benefits of employing an equalizer in a \ac{CAP}-based modulation \ac{WDM} \ac{VLC} system.
A throughput of 8 Gb/s over 1~m indoor link was achieved, with a \ac{FEC} limit of 3.8$\times$10\textsuperscript{-3}. 
In \citep{6881641}, the authors experimentally demonstrated achieved transmissions of 1.35 Gbit/s over a multi-user access \ac{VLC} system based on \ac{WDM} and \ac{RGB} \ac{LED}, by employing a weighted pre-equalization method at the transmitter and a \ac{CMMA} at the receiver 
to compensate the \ac{LED} response.
 
Other efficient techniques such as \ac{MIMO} have been adopted in \ac{VLC} systems for throughput improvement. 
In \citep{7776773}, Hsu \textit{et al.} studied a $3\times 3$ \ac{MIMO} \ac{VLC} system using \ac{OFDM}. The proposed system showed an improvement over the original commercial phosphor white light \ac{LED} (1~MHz), achieving a throughput of 1 Gbit/s over 1~m link reach. 
Hong \textit{et al.} examined the performance of an adaptive $4\times 4$ indoor \ac{MIMO}-\ac{OFDM} \ac{VLC} system in terms of \ac{BER} for different polar angles \citep{7378876}. 
In \citep{de2020rgb}, it was also reported that \ac{WDM} \ac{VLC} systems achieved high data rates. Specifically, \ac{RGB}-based \ac{5G} \ac{VLC} system assisted by a \ac{DPD} attained a throughput of 1.92 Gbit/s, while an \ac{RGB}-based \ac{VLC} system with a hybrid time-frequency domain equalization achieved a throughput of 4.05-Gb/s \citep{zhang20174}. 
Yeh \textit{et al.} examined a $4\times 4$ color polarization-multiplexing method to achieve \ac{LED} bandwidths higher than 465~MHz, which results in 1.7 to 2.3 Gbit/s over 4-m links \citep{8750889}. 
Chvojka \textit{et al.} proposed a polarization division multiplexing technique for \ac{VLC} systems and noted a performance improvement of double the total throughput \citep{chvojka2020visible}. 
 
Another concern is the photodetector area, as it directly impacts the received optical power and, consequently the \ac{SNR}. 
A large photodetector can be used to improve the total power and \ac{SNR}, and thereby increase, the capacitance, which is directly proportional to the photodetector area, limiting the receiver bandwidth and stability \citep{alsulami2018optical}. 

In \citep{8702692}, the authors investigated a \ac{TIA} that improves the tolerance of ultra-high \ac{PD} capacitances, demonstrating a considerable bandwidth enhancement. 
In addition, Nabavi and Yuksel proposed a photodetector array to compose a large reception area, with the aim of reducing the high capacitance and increasing the total received power \citep{9063474}. 
Subsequently, they used a multi-photodetector array prototype together with an amplifier chain in a white phosphorous \ac{LED}, that resulted in a high-throughput performance.

As previously mentioned, optical power levels at the input of the photodetector have an impact the \ac{SNR} and thus, the throughput of the overall \ac{VLC} system. 
Due to limitations on the eye and skin safety specifications, it is imperative that \ac{VLC}  systems avoid high optical powers \citep{OWreport,lopes2020non}. 
The impairments depend on the beam wavelength, power, eye and skin exposure time, and distance. In any case, a \ac{VLC} system must operate under the conditions of the \ac{MPE}, which considers the time the eyes have been exposed to the emission and the maximum allowed power. 
Although infrared communications do not reach the retina, \ac{OWC} systems operating with optical power higher than 10~dBm and beam diameter lower than 1~cm can be hazardous to the eyes \citep{OWreport}. 
On the other hand, \ac{MPE} is very low for \ac{VLC}, since the the eye focuses the visible light 100,000 or even more into the retina \citep{barry1991high}.

\subsection{Uplink Transmission}

In an indoor \ac{VLC} environment, the focus lies on the use of white \acp{LED} to provide simultaneous lighting and communication, typically in downlink transmissions. 
Given the importance of uplink transmissions, it is necessary to investigate such transmissions.
A bidirectional communication in \ac{VLC} is a challenging task \citep{alsulami2018optical}.
The deployment uplink transmissions in \ac{VLC} is generally not viable, since the connected devices might have multiple \acp{LED} in random locations, leading to high costs and may cause some eye discomfort \citep{matheus2019visible}. 
Despite these drawbacks, Umer and Riaz proposed a bidirectional system using visible light with high frequencies for the downlink (green, blue, indigo and violet) and low frequencies for the uplink (red, orange and yellow) in \citep{umer435full}. 
In addition, Wang \textit{et al} experimentally studied a bidirectional \ac{WDM} \ac{VLC} system based on commercial \ac{RGB} \ac{LED} and \ac{PLED} \citep{wang2013demonstration}. 
Pre- and post-equalization techniques were used to compensate the \ac{LED} response, attaining 578-Mbit/s on the downlink and 225~Mbit/s on the uplink, and a \ac{BER} below the \ac{FEC} threshold. 

On the other hand, a hybrid approach that includes \ac{VLC} and \ac{RF} technologies can be used in bidirectional \ac{VLC} links to yield the benefits of each technology. 
In this case, the \ac{VLC} can be used on the downlink to provide high throughput, while \ac{RF} or infrared communication is adopted on the uplink. 
In \citep{8680694}, Pan \textit{et al.} considered 
a \ac{3-D} hybrid indoor system based on \ac{VLC} for the downlink and the \ac{NOMA}-assisted \ac{RF} on the uplink. 
An energy harvesting technique was also adopted, in which the receiver harvests energy through \ac{LED} transmissions.
In addition, several studies have investigated the use of \ac{Wi-Fi} in uplink \ac{VLC} \citep{huang2013design,shao2015design}. In \citep{shao2015design}, Shao \textit{et al.} proposed two distinct hybrid \ac{Wi-Fi}-\ac{VLC} systems where the first implementation used \ac{VLC} and \ac{Wi-Fi} in downlink and uplink transmissions, respectively, and the second one considered both the \ac{VLC} and \ac{Wi-Fi} technologies in the downlink transmissions and \ac{mmWave} in the uplink transmissions \citep{7997701}.

Although \ac{RF}-based uplink transmissions are widely adopted, they are not applicable in environments sensitive to \ac{EMI} such as airspace and hospitals \citep{de2020rgb}. 
To circumvent this issue, infrared-based communication has been proposed as a viable solution.
In \citep{alresheedi2017uplink}, the authors proposed an infrared-based technology for uplink communications using a \ac{FABS} in an effort to improve both throughput and security. 
It was shown that a throughput of approximately 2.5~Gbit/s could be achieve under multipath dispersion, transmitter mobility, and noise.
In \citep{8919555}, the authors proposed an infrared-based system using beam steering in 4 scenarios that include angle diversity receiver (ADR), delay spread and \ac{SNR}, and \ac{OOK}.

\subsection{Security}

As a promising technology for \ac{5G} and \ac{ITS}, security and privacy are essential prerequisites that must be addressed.
In indoor environments, \ac{VLC} systems are less prone to eavesdropping threats and security breach since light does not pass through walls and is contained within the closed environment. 
However, this is not the case in outdoor \ac{VLC}-based systems due to the emission of the broadcast beam, similar to current wireless networks \citep{arfaoui2020physical}.
Hence, security remains a pressing issue that has widely been addressed in the existing literature through encryption-based techniques and/or \ac{PLS} \citep{alsulami2018optical,blinowski2019security}.
In the encryption-based approach for \ac{VLC}, \ac{QKD} can be adopted where the transmitted photons carry a secret key to the receiver.
Specifically, the feasibility of wireless \ac{QKD} for indoor scenarios has been investigated in \citep{elmabrok2018wireless}.
In \citep{mousa2017secure}, Mousa \textit{et al.} proposed a secure \ac{MIMO} \ac{VLC} system that uses the \ac{RSA} technique in the \ac{MAC} layer to encrypt the transmitted information.
The authors also studied the system feasibility to control the encrypted cell size based on the application environment.
In \citep{wang2020optical}, Wang \textit{et al.} proposed an optical encryption technique based on \ac{TGI} employing a micro-\ac{LED}.
The transmitted signal is encrypted by a randomized orthogonal secret key, which attains 4 Gbit/s. 
Through practical implementation, the authors found that the proposed system can support up to 40\% occlusion attacks and 80\% of Gaussian noise in an error-free \ac{VLC} system.

On the other hand, \ac{PLS} has recently emerged as a promising candidate to ensure secrecy in \ac{RF}-based wireless networks.
Hence, recent studies have explored the integration of \ac{PLS} in \ac{VLC} systems~\citep{liu2016physical}.
In \citep{arfaoui2020physical}, Arfaoui \textit{et al.} provided a comprehensive and comparative survey of various information-theoretic techniques.
Specifically, the authors investigated the secrecy performance of consider different channel models, network configurations, input distributions, precoding and input signalling (continuous and discrete), spatial modulation and spatial multiplexing schemes, stochastic geometry, \ac{CSI}, user mobility, device orientation, link blockage and real-life measurement-based channel models.
They also highlighted some avenues for future research studies that include both indoor and outdoor channels with user mobility.

For outdoor applications with autonomous and self-driving vehicles, \ac{VLC} must ensure confidentiality and integrity in the information exchanged. 
In \citep{rowan2017securing}, the authors investigated a secure blockchain-based \ac{V2V} communication through a \ac{VLC} link using a \ac{CMOS}-based camera and acoustic (ultrasonic audio) side channel encoding techniques.
In \citep{wang2018secure}, an \ac{OBU} with multi-level security and a \ac{PLS}-based cooperative communication scheme for \ac{VHN} was designed.
The experimental results revealed the feasibility of the proposed scheme by assuming a successful transmission probability and throughput.
The authors in \citep{8690700} studied the \ac{PLS} of a hybrid \ac{RF}-\ac{VLC} system and evaluated its performance in terms of \ac{SC}.

\subsection{Mobility}

\Ac{VLC} systems can support user mobility by providing reliable communication even in \ac{NLoS} and misaligned conditions, mainly for vehicular applications. 
To this end, efficient handover is required, which occurs when a user on the edge moves to a neighbor access point. 
Consequently, several studies have examined the validation of such applications under real conditions \citep{cuailean2017current}. 
In \citep{1025501}, Zhu \textit{et al.} studied various handover mechanisms based on \ac{LED} traffic lights and moving vehicles. 
Okada \textit{et al.} explored a road-to-vehicle \ac{VLC} system based on a transmission between an
\ac{LED} traffic light serving as transmitter and a \ac{PD} to enable \ac{ITS} \citep{5164423}. 
Furthermore, the authors employed two cameras and applied imaging optics as a tracking mechanism to control the alignment between the transmitter and receiver. 
Although the solution proved to be efficient in terms of cost, mobility, and communication range, the system is relatively complex from an implementation perspective. 
It is possible to consider multiple receivers in which case some signal processing techniques can be used to decide which receiver has the highest \ac{OSNR}. 
An approach to adjust the angle between the transmitter and the receiver was investigated using light sensors in \citep{jeong2016receiver}. 
The alignment correction was determined on the basis of the measured received optical power. 
In \citep{cuba2020cooperative}, the authors studied the feasibility of \ac{VLC} between moving cars considering two distinct scenarios that include a multiple-lane rectilinear roadway and a multiple-lane curvilinear roadway. 
The authors explored the feasibility of a full-duplex cooperative communication protocol that minimizes communication disruption in the event of \ac{NLoS} conditions.   

It should be mentioned that autonomous vehicles need to accurately estimate the position of surrounding vehicles and obstacles on the road.
Data exchange must occur at rates higher than 50~Hz and provide centimeter-accuracy to avoid potential accidents in \ac{V2V} or platooning applications. 
In \citep{SonerBurak2019VVLP}, Soner and Coleri proposed a \ac{VLP} solution for \ac{VPE}. 
However, this approach does not require novel \ac{VLC} receivers. 
A quadrant photodetector with low-cost and size, high throughput, and accurate angle-of-arrival sensing, was employed at the receiver.
An estimation algorithm that employs data from two receivers and determines the position of neighbouring cars by means of triangulation, was proposed. 
Finally, simulation and sensitivity analyses were performed on \ac{SUMO}, demonstrating an accurate estimation under realistic channel conditions.    

Considering the \ac{VLC} mobility for indoor scenarios, Burton \textit{et al.} presented the concept, design and analysis of a {VLC} receiver to support mobility and prevent signal disruption in home and office environments \citep{burton2014design}. 
The receiver employs seven sub‐receivers to provide angular and spatial diversity.
The signal with the highest \ac{SNR} is selected by means of selection combining technique. 
The system is then evaluated in terms of received power, impulse response, \ac{SNR} and delay spread \ac{RMS} and compared with one that only has a single receiver. 
In \citep{7109107}, Nuwanpriya \textit{et al.} proposed two novel and practical designs of angle diversity that enable \ac{MIMO} transmissions in indoor \ac{VLC} environments. 
The receivers consist of several photodetectors with different angles in order to reap the benefits of \ac{MIMO} channels without the requirement of spatial separation. 
Through analytical, simulation, and experimental analyses, remarkable BER performance and throughput were achieved to support user mobility in various geographic locations without the need of hardware adjustments at the receivers

\section{Conclusions}
\label{concl}
Due to the attractive features of \ac{VLC}, it has recently emerged as a key component for future short-distance networks with high-data throughput and security. 
Despite its enormous potential, vehicular \ac{VLC} is faced with many challenges (such as the  undesirable effects of sunlight) that may hinder its future development and deployment. 
This tutorial paper covers a wide range of topics associated with the \c{VLC} systems, which includes the basic design of the \ac{VLC} systems, channel models, topologies (such as a peer-to-peer, star, and broadcast), standardization efforts of {VLC} that includes including IEEE 802.15.7 standard. 
The most relevant challenges regarding the practical applications of \ac{VLC} systems are also discussed. These challenges include unwanted effects such as the influence of the sun and artificial light sources.
Some key performance metrics useful to evaluate the underlying systems, such as the throughput, security and mobility, are also discussed.
Besides, this tutorial paper also addresses challenges pertaining to \ac{VLC} system, such as flickering, dimming, and uplink.

\section{Acknowledgments}
This work is partly supported by the Research Council of Finland through ECO-NEWS n.358928 and X-SDEN n. 349965, by Business Finland through 6G REEVA n.10278/31/2022, and by EU MSCA project ``COALESCE'' under Grant Number 101130739.

\bibliographystyle{IEEEtran}
\bibliography{references}

\end{document}